\documentclass[a4paper]{jpconf}
\usepackage{graphicx}
\usepackage{mathtools}
\usepackage{physics}

\begin{document}

\title[Featuring causal order in teleportation of two quantum teleportation channels]{Featuring causal order in teleportation of two quantum teleportation channels}


\author{Carlos Cardoso-Isidoro and Francisco Delgado}

\address{{Tecnologico de Monterrey, Engineering and Sciences School, 
M\'exico.}}

\ead{fdelgado@tec.mx}

\begin{abstract}
Causal order can improve the quantum information transmission in teleportation due a noisy entangled resource by using an appropriate measurement state on the control state ruling the causal order. In this work, we get analytically the fidelity for the entire process under an arbitrary measurement over such control state and then we get a perfect teleportation by selecting the optimal measurement on it. We also analyse other values characterizing the imperfect entangled state where a perfect teleportation can not be reached. Notably, we determine that the best fidelity does not depend on the probabilities of the control state but on the imperfect initial resource entangled.
\end{abstract}

\section{Introduction}\label{sec1}

Quantum communication is always looking for improvements. Recently, it has been shown, that a couple of depolarizing channels becomes transparent as a result of interference under indefinite causal order reaching enhancement with the assistance of causal order activation \cite{ebler1}, thus it is widely studied. It is the case for the so called quantum switch \cite{Chiribella1}. 

Causal order enhancements in communications have had approaches both theoretical and experimental. In the case of two quantum channels, it has been shown that even though no information can be transmitted through depolarizing channels by classical means due to the noise, it could be possible to transmit information by combining two depolarizing channels in a superposition of causal orders \cite{ebler1, Goswami1}. Thus, it has been extended to more than two channels, showing the enhancement of transmission for the three channel scenario \cite{Procopio1} improves the amount of information transmitted. Alternatively, for the quantum teleportation algorithm with very noisy singlets introducing serious imperfections in the process, it has been shown the possibility of still transmitting perfectly an arbitrary information state by applying superposition of causal order of two such channels \cite{Chiranjib1}. Such work states that for an egalitarian superposition on the qubit controlling the causal order, certain final measurements on it let to reach a perfect teleportation for the nosiest entangled resource using a two teleportation channels scheme.

This work shows that a perfect quantum teleportation can be still reached using causal order superposition with a proper selection of the post-measurement state of a general control system, thus extending the outcomes obtained in \cite{Chiranjib1}. Section \ref{sec2} presents details of the teleportation circuit in causal order for two channels. Section \ref{sec3} shows the main findings for optimal measurements on the control through the corresponding fidelity, analyzing the conditions for which an optimal teleportation can be obtained. The final section states the conclusions. 

\section{Formalism to set quantum teleportation under an indefinite causal order scheme with two channels}\label{sec2}

The aim of this work is the teleportation of the state $\left|\psi \right> = \cos{\frac{\theta_0}{2}}\left|0 \right>+\sin{\frac{\theta_0}{2}}e^{i\phi_0}\vert1\rangle$. In \cite{Chiranjib1} the proposed imperfect entangled state assessing the teleportation is $\left|\chi \right>=\sum_{i=0}^{3}p_i\left|\beta_i \right>$; where $\left|\beta_i\right>$ are the Bell states $\left|\beta_0 \right>=\left|\beta_{00} \right>$, $\left|\beta_1 \right>=\left|\beta_{01} \right>$, $\left|\beta_2 \right>=\left|\beta_{11} \right>$ and $\left|\beta_3 \right>=\left|\beta_{10} \right>$, with:

\begin{equation}
    \left|\beta_{ij}\right>=\frac{1}{\sqrt{2}}\left(\left|0 \hspace{0.2cm} j \right>\right)+(-1)^{i}\left|1 \hspace{0.2cm} j\oplus 1 \right>
\end{equation}

The perfect teleportation process could be achieved with $p_0=1$, ($p_1=p_2=p_3=0$). If the Bell state $\left|\beta_{0} \right>$ is pretended to be used as the successful entanglement resource, the output of the channel is given by $\Lambda[\rho]=\sum_{i=0}^3p_i\widetilde{\sigma}_i\rho\widetilde{\sigma}^\dagger_i=\sum_{i=0}^3p_i\sigma_i\rho\sigma_i$ \cite{Bowen1}, with $\widetilde{\sigma}_i=\sigma_i$ if $i=0,1,3$ and $\widetilde{\sigma}_2=i\sigma_2$, where $\rho=\left| \psi\right> \left< \psi\right|$. Thus, for a single teleportation channel, the corresponding Kraus operators are $K_i=\sqrt{p_i}\sigma_i$. Recently, \cite{Chiranjib1} presents a causal order version of two teleportation channels in superposition of causal orders for two teleportation channels controlled by the state:
\begin{equation}\label{rhoc}
    \rho_c=\left|\psi_c \right>\left<\psi_c \right|= \sum_{ k_i k'=0}^1 \sqrt{q_k q_{k'}} \vert k^{} \rangle \langle k' \vert
\end{equation}

\begin{figure}[b]
    \centering
    \includegraphics[scale=0.5]{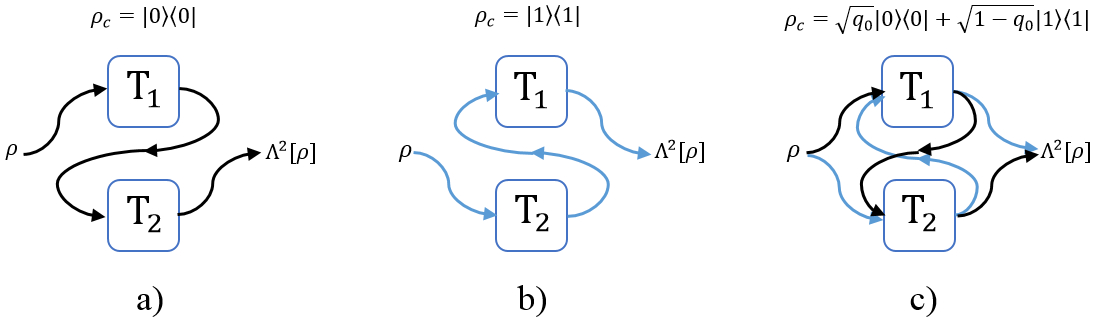}
    \caption{Two teleportation channels in a) - b) a definite causal order as function of control states $\vert 0 \rangle \langle 0 \vert$ or $\vert 1 \rangle \langle 1 \vert$, and c) an indefinite causal order as a superposition of previous cases.}
    \label{fig1}
\end{figure}

Figure \ref{fig1} shows the deployment of two teleportation channels a) - b) in a definite causal order as function of the control state, ${\rm T}_1$ first and then ${\rm T}_2$ if $\vert \psi_c \rangle=\vert 0 \rangle$  or ${\rm T}_2$ first and then ${\rm T}_1$ if $\vert \psi_c \rangle=\vert 1 \rangle$; and c) in an indefinite causal order if $\vert \psi_c \rangle$ is a superposition of the two last control states with probabilities $q_0$ and $q_1=1-q_0$ respectively as in (\ref{rhoc}). 

In such process, a final measurement on the control state shows when the outcome is $\left|+\right>=\frac{1}{\sqrt{2}}(\vert 0 \rangle + \vert 1 \rangle) $, a perfect teleportation process is obtained for the worst case given by $p_1=p_2=p_3=p=\frac{1}{3}$. In this work,  we consider instead the general state for the measurement as (which could exhibit different probabilities of success ${\mathcal P}$):
\begin{equation}
    \left|\psi_{ m} \right> = \cos{\frac{\theta}{2}}\left|0\right> + \sin{\frac{\theta}{2}} e^{i\phi}\left|1\right> \label{MesSt} 
\end{equation}

Following \cite{Chiranjib1} we construct the Kraus operators for two consecutive teleportation channels in causal order, but instead of apply a measurement on the control state based on the states $\vert \pm \rangle$, we apply a general measurement based on the state (\ref{MesSt}) and their correspondent orthogonal state $\left|\psi^{\perp}_{ m} \right> = \sin{\frac{\theta}{2}}\left|0\right> - \cos{\frac{\theta}{2}} e^{-i\phi}\left|1\right>$. For the first state, we get the unnormalized output:
\begin{equation}
    \Lambda_{\rm un}^2[\rho]=\sum_{i,j=0}^{3}p_ip_j\left((\frac{1}{2}+(q_0-\frac{1}{2})\cos{\theta}) \sigma_i\sigma_j\rho \sigma_j\sigma_i + \sqrt{q_0q_1}\sin{\theta}\cos{\phi} \sigma_i\sigma_j\rho \sigma_i\sigma_j \right)
\end{equation}

\section{Fidelity of quantum teleportation under an indefinite causal order scheme}\label{sec3}

We use the fidelity $\mathcal{F}_{\rm} =\rm{Tr}\left(\Lambda^2[\rho]\rho \right)$ to assess the entire process (a comparative measure between the input and the teleported states) together with the measurement probability $\mathcal{P}_{\rm} =\rm{Tr}\left(\Lambda^2[\rho] \right)$:
\begin{eqnarray}
    \mathcal{F}_{\rm un} &=& \sum_{i,j=0}^3p_ip_j\left((\frac{1}{2}+(q_0-\frac{1}{2})\cos{\theta}) {\rm Tr}(\rho \sigma_i\sigma_j\rho\sigma_j\sigma_i) +\sqrt{q_0q_1}\sin{\theta}\cos{\phi}\rm{Tr}(\rho \sigma_i\sigma_j\rho\sigma_i\sigma_j) \right) \label{UnFidelity} \\
    {\mathcal P} &=& \sum_{i,j=0}^3p_ip_j\left((\frac{1}{2}+(q_0-\frac{1}{2})\cos{\theta}){\rm Tr}( \sigma_i\sigma_j\rho\sigma_j\sigma_i) +\sqrt{q_0q_1}\sin{\theta}\cos{\phi}{\rm Tr}( \sigma_i\sigma_j\rho\sigma_i\sigma_j) \right) \label{Probability}
\end{eqnarray}

\noindent then, dividing (\ref{UnFidelity}) by (\ref{Probability}), we get the normalized fidelity:

\begin{equation}\label{Fidelity}
    \mathcal{F} =\frac{ \sum_{i,j=0}^3p_ip_j\left((\frac{1}{2}+(q_0-\frac{1}{2})\cos{\theta}){\rm Tr}(\rho \sigma_i\sigma_j\rho\sigma_j\sigma_i) +\sqrt{q_0q_1}\sin{\theta}\cos{\phi}{\rm Tr}(\rho \sigma_i\sigma_j\rho\sigma_i\sigma_j) \right)}{\sum_{i,j=0}^{3}p_ip_j\left((\frac{1}{2}+(q_0-\frac{1}{2})\cos{\theta}) {\rm Tr}(\sigma_i\sigma_j\rho \sigma_j\sigma_i) + \sqrt{q_0q_1}\sin{\theta}\cos{\phi} {\rm Tr}(\sigma_i\sigma_j\rho \sigma_i\sigma_j) \right)} 
\end{equation}

\begin{figure}[ht]
    \centering
    \includegraphics[scale=1]{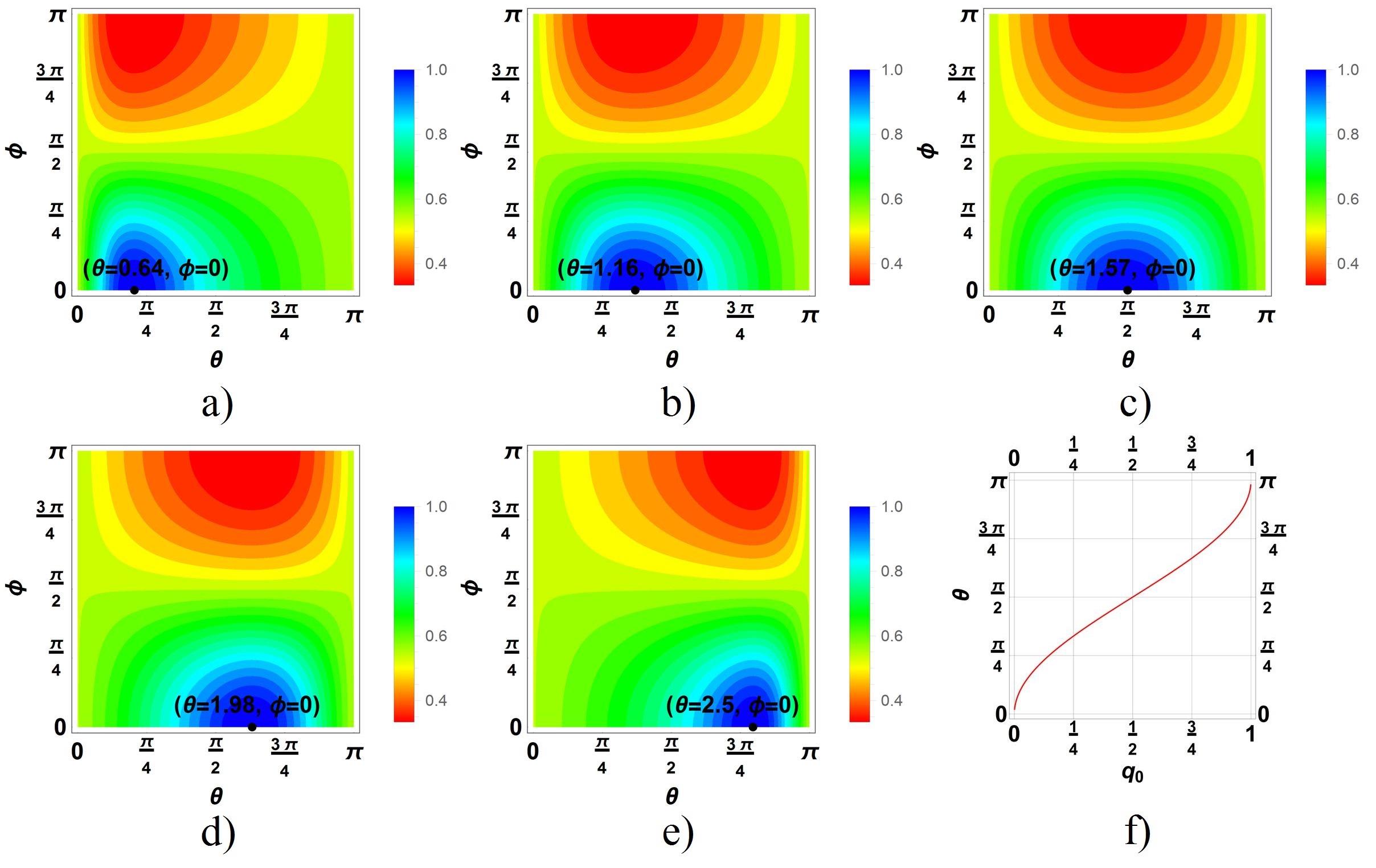}
    \caption{In the contour plots a) - e) for ${\mathcal F}$, it is indicated the values for $\theta$ and $\phi$ such that $\mathcal{F}=1$ is reached ($q_0=0.1, 0.3, 0.5, 0.7, 0.9$ and ${\mathcal P}=0.12,0.28,0.33,0.28,0.12$ respectively). Color bar shows the values of fidelity. Plot f) exhibits the relation between $\theta$ and $q_0$ under the election of the best control measurement ($\phi=0$ always).}
    \label{fig2}
\end{figure}

We can still simplify last formula using the Pauli matrices and the trace operation properties: ${\rm Tr}(\sigma_i\sigma_j\rho \sigma_j\sigma_i)=1$ and ${\rm Tr}(\sigma_i\sigma_j\rho \sigma_i\sigma_j)=\delta_{ij}+(1-\delta_{ij}){\rm sgn}(ij)$, with ${\rm sgn}(x)$ the sign function. Now, our task is to demonstrate for the worst case $p_1=p_2=p_3=p=1/3$ and $p_0=0$ that it is possible to reach a perfect teleportation ($\mathcal{F}=1$) by choosing adequately the measurement state (\ref{MesSt}). We solve numerically the optimization problem by fixing $q_0$, the last $p_i$ values and then finding the best $\theta$ and $\phi$ in (\ref{MesSt})  maximixing $\mathcal F$ in (\ref{Fidelity}). 

Figure \ref{fig2} shows some illustrative graphs where $\mathcal{F}=1$ is reached by varying the measurement state. Contour plots a) - e) correspond to $q_0=0.1, 0.3, 0.5, 0.7$ and $0.9$ deploying the fidelity values for the entire measurement states ($\ref{MesSt}$) in the Bloch representation in agreement with the color bar besides. The best state is marked with a black dot stating the corresponding values of $\theta$ and $\phi$. The case in \cite{Chiranjib1} has the best ${\mathcal P}=0.33$. Plot \ref{fig2}f shows $\theta$ versus $q_0$ (note $\phi=0$ in all cases). Note \ref{fig2}c corresponds to the case analyzed in \cite{Chiranjib1} where $\vert \psi_c \rangle =\left|+\right>$ as optimal measurement. 

\begin{figure}
    \centering
    \includegraphics[scale=1.6]{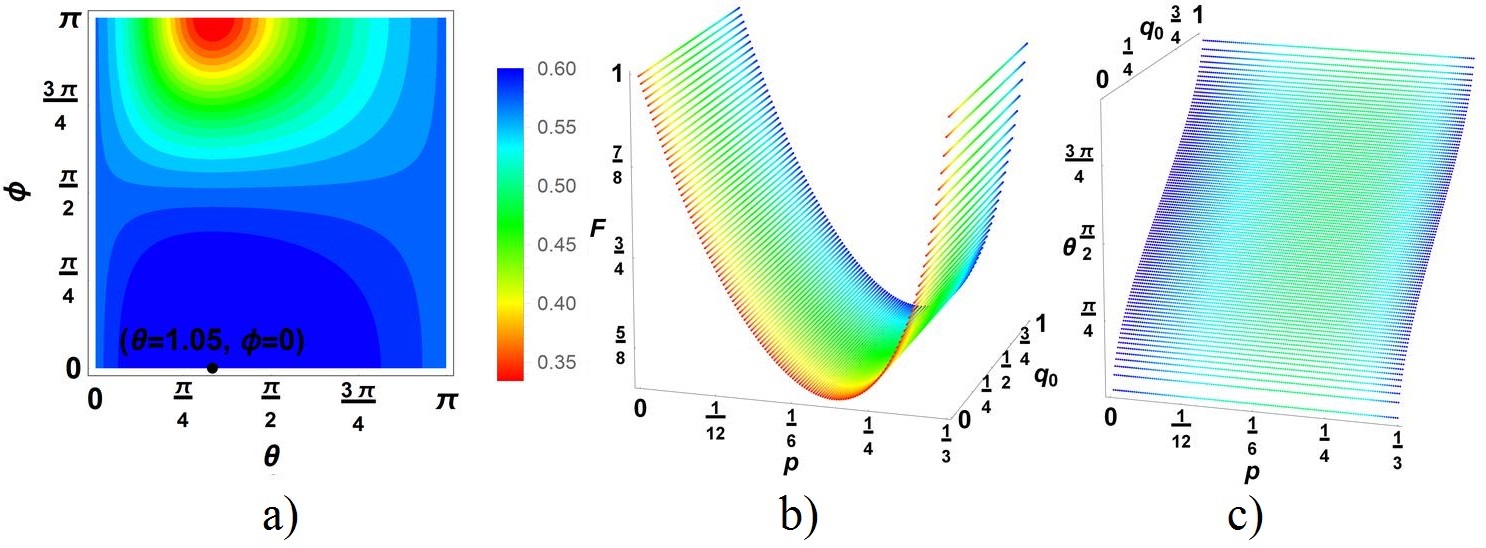}
    \caption{Outcomes for the fidelity ${\mathcal F}$ for other values of $p$ different from $\frac{1}{3}$. a) Shows the contour plot for ${\mathcal F}$ for $p=\frac{1}{6}$ and $q_0=\frac{1}{4}$; b)-c) depicts the dependence of  ${\mathcal F}$ and $\theta$ from $p$ and $q_0$.}
    \label{fig3}
\end{figure}

We can ask about the best performance for cases $p \ne \frac{1}{3}$. Unfortunately, the situation is not optimal there. Plot \ref{fig3}a shows the best fidelity (${\mathcal F}=0.60, {\mathcal P}=0.28$) for $p=\frac{1}{6}$ and $q_0=\frac{1}{4}$. In fact, surprisingly the outcome is independent from $q_0$: once selected the $p$-value, the best fidelity becomes fixed. Thus, Figure \ref{fig3}b reproduces the curve reported in Figure 2 of \cite{Chiranjib1} for the $q_0=\frac{1}{2}$ case. Finally, we report the dependence of $\theta$ with $p$ and $q_0$ (\ref{fig3}b), the same than in Figure \ref{fig2}f.

\section{Conclusions}\label{sec4}
We have analyzed the features of two teleportation channels in superposition. For the specific case with $p=1/3$, we have shown that $\mathcal{F}=1$ can be reached not only with the state $\vert + \rangle$ for $q_0=\frac{1}{2}$, but as far as we choose the measurement state given by (\ref{MesSt}) in the range for $\phi=0$ and $\theta \in [0,\pi]$, a perfect teleportation can be done as function of arbitrary $p_0$. If we analyze other values for $p\neq 1/3$, we can not reach $\mathcal{F}=1$ no matter the measurement state chosen. Nevertheless, we can figure out the best choice for the measurement state in order to reach the optimal fidelity, noting it does not depend on the value for $q_0$, so that the maximum for $\mathcal{F}$ is fixed once selected $p$ and it can be reached with the correct values for $\theta$ and $\phi$ in the measurement basis. Future work should to analyze similar outcomes for $\mathcal{F}$ by managing independently the $p_0,p_1,p_2,p_3,q_0$ values together with $\theta$ and $\phi$ in order to search the maximum $\mathcal{F}$ for each case.

\end{document}